# Hole spin in tunable Ge hut wire double quantum dot

Gang Xu,[1,2,] Fei Gao,[3] Ke Wang,[1,2] Ting Zhang,[1,2] He Liu,[1,2] Gang Cao,[1,2] Ting Wang,[3] Jian-Jun Zhang,[3] Hong-Wen Jiang,[4] Hai-Ou Li,[1,2*] and Guo-Ping Guo[1,2,5*]

[1] *CAS Key Laboratory of Quantum Information, University of Science and Technology of China, Hefei, Anhui 230026, China*

[2] *CAS Center for Excellence and Synergetic Innovation Center in Quantum Information and Quantum Physics, University of Science and Technology of China, Hefei, Anhui 230026, China*

[3] *Institute of Physics and CAS Center for Excellence in Topological Quantum Computation, Chinese Academy of Sciences, Beijing 100190, China*

[4] *Department of Physics and Astronomy, University of California, Los Angeles, California 90095, USA*

[5] *Origin Quantum Computing Company Limited, Hefei, Anhui 230026, China*

[*]Emails: haiouli@ustc.edu.cn; gpguo@ustc.edu.cn

**Abstract:** Holes in germanium (Ge) exhibit strong spin-orbit interaction, which can be exploited for fast and all-electrical manipulation of spin states. Here, we report transport experiments in a tunable Ge hut wire hole double quantum dot. We observe the signatures of Pauli spin blockade (PSB) with a large singlet-triplet energy splitting of ~1.1 meV and extract the g factor. By analyzing the the PSB leakage current, we obtain a spin-orbit length $l_{so}$ of ~ 40–100 nm. Furthermore, we demonstrate the electric dipole spin resonance. These results lay a solid foundation for implementing high quality tunable hole spin-orbit qubits.

For spin-based quantum computing, the long coherence time and fast electrical controllability of the spin states are two key characteristics.[1] Therefore, searching for suitable materials has become significant over the past few years. The group IV semiconductors, such as silicon (Si) and germanium (Ge), have predominantly stable isotopes with zero nuclear spin, which limits the hyperfine interaction, leading to a long dephasing time.[2-6] Holes in Ge have an even weaker hyperfine interaction and a stronger spin-orbit interaction (SOI), which can be exploited for faster spin manipulation. Therefore, the Ge quantum dots is an ideal platform for quantum computing. [7, 8]

Over the past few years, Ge quantum dots have been investigated extensively in Si/Ge core-shell nanowire systems, [9-12] but the cylindrical geometry of the Si/Ge core-shell nanowires leads to a mixture of heavy holes (HH) and light holes (LH), which results in a non-Ising-type hyperfine interaction and negatively affects the coherence time.[3] Si/Ge islands device has been demonstrated by means of Stranski-Krastanow (SK) growth mode.[13] However, due to the small size of the Si/Ge islands, it is difficult to create the double quantum dot (DQD) structures typically used in spin manipulation experiments. In 2012, Ge hut wire growth by molecular beam epitaxy is implemented.[14] The Ge hut clusters can expand into Ge hut wires under specific conditions, with length exceeding one micrometer. Previous experiments based on Ge hut wires include hole transport in a single quantum dot, hole-resonator coupling and a hole spin qubit.[15-18] However, the first hole spin qubit based on a Ge hut wire was realized in a DQD with only two plunger gates, thus the interdot tunneling is not fully tunable, [18] which impeded the further research on two-qubit control. Systematic investigations of the hole spin states in a tunable DQD not only improve the understanding of spin relaxation mechanisms in Ge hut wires but also provide guidance on better spin qubit control.

Figure 1 (a) shows an atomic force microscopy image of the Ge hut wires epitxially grown by the means of the SK growth mode. A scanning electron microscopy image of a Ge hut wire DQD device is shown in Fig. 1 (b). The patterning of the electrodes was performed by electron beam lithography. After removing the native oxide of the Si capping layer in buffered hydrofluoric acid, 30-nm-thick Pd contacts were deposited as the source and drain leads of a DQD. Then, a 30-nm-thick alumina layer was grown by the atomic layer deposition. Finally, five 3/25-nm-thick Ti/Pd top gates were deposited, with each gate being 30 nm wide. The multilayer gate structures are used to induce tunneling barriers to define the DQD. Gates G1 and G5 are applied to set the outer barrier, while gates G2 and G4 are used to tune the electrochemical potential of each dot separately. The middle gate G3 is used to control the interdot tunnel coupling strength between the left dot and right dot.

The measurements were performed at approximately 240 mK in a liquid He-3 refrigerator. The quantum transport signal through the source/drain leads is measured by a multimeter after passing through a low-noise preamplifier. To show that the interdot tunnel coupling can be tuned by the middle gate voltage $V_{G3}$, charge stability diagrams at different values of $V_{G3}$ are presented in Figs. 1(c)-(e). We keep the outer barrier gates at constant voltages ($V_{G1} = -0.82$ V and $V_{G5} = -0.2$ V), and the current is plotted as a function of $V_{G2}$ and $V_{G4}$ at a fixed value of $V_{SD} = 0.5$ mV. At $V_{G3} = 0$ V (Fig. 1c), the interdot tunnel coupling is weak, as suggested by the isolated triple points. Making $V_{G3}$ more negative increases the interdot tunnel coupling, and we observe only faintly visible long edges of the honeycombs [$V_{G3} = -0.02$ V, Fig. 1(d)]. Finally, at $V_{G3} = -0.04$ V [Fig. 1(e)], the interdot tunnel coupling becomes strong. We obtain the characteristic honeycomb pattern for the charge stability diagram of the DQD. In Fig. 1(f), we show a stability diagram of a DQD weakly coupled to the reservoirs at $V_{SD} = 2$ mV. A very regular pattern of bias triangle pairs is clearly visible. From the dimensions of the honeycombs and bias triangles, the charging energies and lever arms of the system can be extracted as $E_C^L = 4.3$ meV ($E_C^R = 4.6$ meV) and $\alpha_L = 0.14$ ($\alpha_R = 0.13$).

Pauli spin blockade (PSB) [19, 20] is a key technique for implementing spin-to-charge conversion and directly investigating spin relaxation mechanisms in DQD systems. The PSB originates from spin-conserved tunneling during the interdot transition. For example, the cycle (1, 0) → (2, 0) → (1, 1) → (1, 0) transfers one hole from left to right [Fig. 2(a)]. However, at the opposite bias $V_{SD}$, the transition $(1, 1) \rightarrow (2, 0)$ is forbidden when the (1, 1) state is a triplet and the accessible (2, 0) state is a singlet [Fig. 2(b)].

By comparing the data for the negative and positive $V_{SD}$ in Figs. 2(a) and 2(b), we observe current rectification with suppressed transport only for positive $V_{SD}$. From the measurements shown in Fig. 2(b), the energy gap between the baseline and the high current line marked by the green dashed line is the singlet-triplet energy splitting $\Delta_{ST}$. It can be found that $\Delta_{ST}$ in the DQD is 1.1 meV in this gate voltage region. Figure 2(c) shows the transport characteristics of the DQD in the triangle region with the same gate voltages as in Fig. 2(b) but for a magnetic field of B = 2 T applied perpendicular to the device. In this finite magnetic field, the T (2, 0) triplet state splits into three states: $T_-(2, 0)$, $T_0(2, 0)$ and $T_+(2, 0)$. Now, $\Delta_{ST}$ is measured by the energy gap between the S (2, 0) and the $T_-(2, 0)$ states. Thus, $\Delta_{ST}$ is smaller when compared with the value at zero magnetic field due to Zeeman splitting. Figure 2(d) shows $\Delta_{ST}$ as a function of the applied magnetic field B. It can be seen that with increasing B, $\Delta_{ST}$ decreases linearly following the equation $\Delta_{ST}(B) = \Delta_{ST}(0) - g\mu_B B$, where $\mu_B$ is the Bohr magneton and g is the effective Landé g factor. The measurement of spin blockade was not performed in the few-hole region, a larger occupation number may change the effective aspect ratios of the DQD, thus, increase its HH–LH mixing. The ground states of the Ge hut wire DQD are almost HH and admixed with LH state.[15] A linear fit yields g = 3.4 ± 0.2, which is consistent with previous reports.[15, 16]

More specifically, we focus on the magnetic field evolution of the leakage current in the PSB region. The leakage current can result from different spin relaxation mechanisms, such as spin-flip cotunneling and the SOI.[21-26] Previous studies have shown that the zero-field dip in the leakage current that occurs in a DQD signifies a strong SOI.[24, 25] The dip is usually explained in terms of a competition between different types of spin-mixing processes: the combination of the SOI and Zeeman splitting due to the applied magnetic field enables transitions between triplet and singlet configurations. This mechanism becomes more efficient at a higher magnetic field and thus produces a dip in the leakage current around zero magnetic field.

Figure 3(a) presents the typical spin blockade region, Fig. 3(c) maps the leakage current in the spin blockade region as a function of the out-of-plane magnetic field B and the detuning axis [Fig. 3(a), white arrow)]. Figure 3(e) shows a line cut around $\varepsilon = 0$ in Fig. 3(c), revealing a double-peak structure with a dip at zero magnetic field. The suppression of the leakage current at B = 0 and the rapid lifting of the spin blockade with a finite field signify strong SOI-induced hybridization of the T (1, 1) and S (2, 0) states. The data are fitted using theory from[26]

$$I(B) = \Gamma_{rel} \frac{[\omega - B^2 + \tau^2][\omega(1+4\gamma) + B^2 - \tau^2]}{6\gamma\omega^2 + 2B^2\eta^2 t^2}, \tag{1}$$

where $\Gamma_{rel}$ is the spin relaxation rate, $\omega = \sqrt{(B^2 - \tau^2)^2 + 8B^2\beta^2 t^2}$, $\gamma = \Gamma_{rel}/\Gamma$, $\Gamma$ is the decay rate to lead, $\tau = t\sqrt{1 + 3\beta^2}$, $t$ is the interdot coupling strength, $t_{so}$ is the spin-orbit coupling strength and $\eta = t_{so}/t$. From the fitting, we can extract $t$ ~45 ± 5 μeV and $t_{so}$ ~ 27 ± 3 μeV.

To further study the influence of the SOI on spin relaxation, we introduce more holes into the DQD by tuning $V_{G2}$ and $V_{G4}$ more negative. We observe a similar phenomenon in another PSB region shown in Fig. 3(b), by measuring the leakage current as a function of the magnetic field B and the detuning axis [Fig. 3(d)]. We also find a double-peak structure with a dip at zero magnetic field but with a larger width of the overall double-peak structure shown in Fig. 3(f). The interdot coupling strength and spin-orbit coupling strength are extracted as $t$ ~ 84 ± 7 μeV and $t_{so}$ ~ 38 ± 4 μeV respectively. The stronger $t_{so}$ in this charge configuration may be due to the enhancement of the spin state mixing by the SOI, which is related to the dot size and the orbital level. In addition, the suppression of the current at the high magnetic fields could indicate that B exceeds the effective level width of S (2, 0) by such an amount that the system is pushed into a Coulomb blockade in the lowest-lying (1, 1) triplet state.

The spin-orbit length $l_{so}$ is also a direct measure of the SOI strength: a stronger SOI results in a shorter $l_{so}$. Here we simply parameterize this with $t_{so}/t \sim l/ l_{so}$,[22] the size of a Ge hut wire QD $l$ is calculated to be $l=\hbar/\sqrt{(m^*\Delta_{ST})}$~20–50 nm, where the effective mass is between the LH and the HH effective masses in Ge hut wire, $m_{LH} < m^* < m_{HH}$, with $m_{LH} = 0.042 m_e$ and $m_{HH} = 0.32 m_e$[8], where $m_e$ is the free electron mass. The S-T orbital level spacing $\Delta_{ST}$ is approximately 1.1 meV. Thus, the spin-orbit length $l_{so}$ is estimated to be ~ 40–100 nm. The $l_{so}$ obtained in this study is shorter than the values reported in InAs ~100 nm and InSb ~200 nm[27, 28] and Ge/Si core-shell nanowires ~100 nm[26]. The shorter $l_{so}$ signifies a hole moving through the hut wire undergoes spin precession around with a π rotation over a shorter distance. Which implies a stronger SOI in Ge hut wire system and faster electrical controllability of the spin states. These results are in good agreement with the assumption of a strong spin-orbit coupling, implying the possibility of performing EDSR and achieving a spin-orbit qubit.[18, 29]

Since spin blockade can be lifted by the single-spin rotations. The Ge hut wire cross-section does not have a center of inversion symmetry, the SOI in our system is Rashba type in the presence of an out-of-plane electric field.[8, 15] In the presence of strong SOI in Ge hut nanowire, spin rotations can be induced by the applied microwave-frequency electric field combination with a constant magnetic field. The EDSR mechanism is expected to drive single hole spin rotations whenever the frequency of electric field matches the Zeeman energy of the hole spin $f_0 = g\mu_B B/h$, where g is the Landé g-factor, $\mu_B$ is the Bohr magneton, B is the applied static magnetic field, and h is the Planck's constant. We set the DQD in PSB region as shown in Fig.3 (a), and apply a microwave-frequency electric field to plunger gate G4 at finite magnetic field B. The spin-orbit state of the double dot rapidly changes from parallel to antiparallel, the blockade is lifted. To detect the EDSR, we measure the leakage current as a function of magnetic field and microwave frequency [Fig. 4(a)]. From the EDSR spectrum line, we extracted the hole Landé g-factor ~ 3.5, which is similar to the results of Fig.2 (second DQD device).

At fixed frequency of the microwave, the line width of the EDSR peak also allows the extraction of a lower limit for the hole spin dephasing time $T_2^*$.[18] As shown in Fig. 4(b), the dephasing time of about $8 \pm 2$ ns can be extracted using the relation $T_2^* = 2\hbar\sqrt{\ln 2}/g\mu_B \Delta B_{EDSR}$,[18, 30] where $\Delta B_{EDSR}$ is the full width at half maximum of the resonance peak ~ $1.5 \pm 0.3$ mT. The shorter $T_2^*$ compared with reported in ref 18 which out-of-plane magnetic field dephasing time is estimated of $T_2^*$~33 ns may be attributed to two reasons. Firstly, the $\Delta B_{\text{EDSR}}$ is broadened due to a higher measured temperature, [31] the base temperature in our study is ~240 mK and in ref 18 is ~40 mK. Secondly our measurements were performed at a relatively high RF power, the smaller peak width was not obtained. [18] But this value gives a lower bound for the inhomogeneously broadened spin dephasing time, which is significantly for efficient spin manipulation.

In conclusion, we have presented transport measurements in a tunable Ge hut wire hole DQD. We observe a large singlet-triplet energy splitting of ~1.1 meV and extract the effective Landé g factor. In addition, the PSB under different charge configurations is also identified. Furthermore, we illustrate the importance of the SOI for controlling the hole spin and obtain a shorter spin-orbit length of ~40–100 nm in our system. Finally, we demonstrate the electrical control of single hole spin rotations by using EDSR. These findings offer new insights into the properties of the hole spin and spin-orbit-induced quantum states mixing[28, 32] in Ge hut wires, and also underline the potential of holes in Ge hut wire has long lived electrically tunable spin qubits.

**Acknowledgements:**
This work was supported by the National Key Research and Development Program of China (Grant No.2016YFA0301700), the National Natural Science Foundation of China (Grants No. 61674132, 11674300, 11625419 and 11574356), the Strategic Priority Research Program of the CAS (Grant Nos. XDB24030601 and XDB30000000), the Anhui initiative in Quantum information Technologies (Grants No. AHY080000) and this work was partially carried out at the USTC Center for Micro and Nanoscale Research and Fabrication.

**Figure captions:**

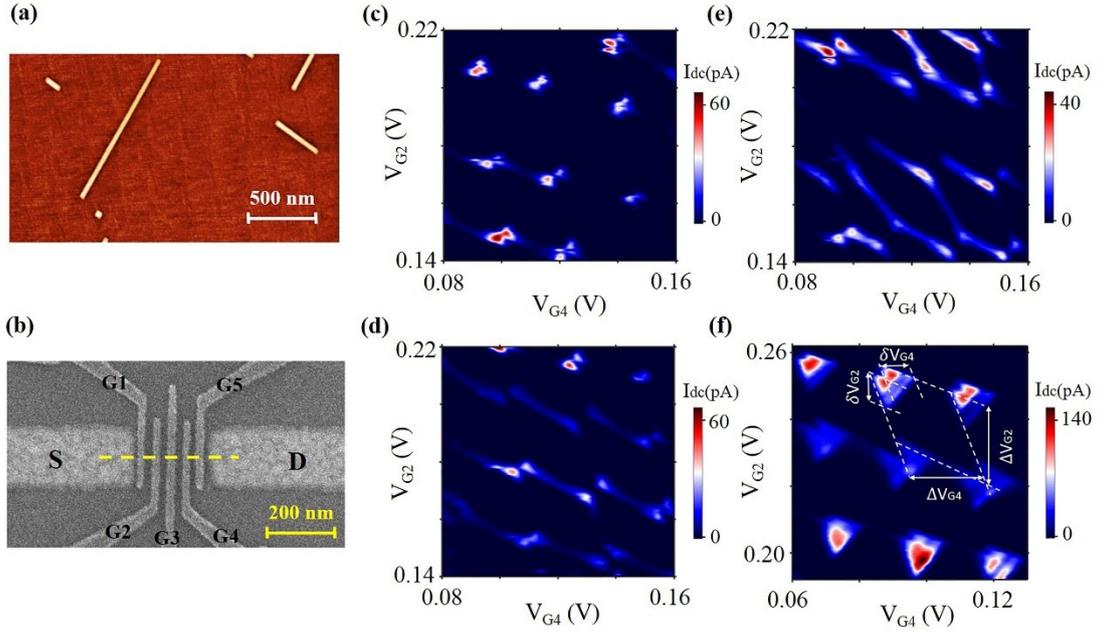

**Fig. 1.** (a) An atomic force microscopy image of Ge hut wires grown on a Si substrate. (b) The scanning electron microscopy image of a Ge hut wire (yellow dashed line) DQD device. (c)-(e) Source-drain current as a function of $V_{G2}$ and $V_{G4}$ for various gate voltages $V_{G3}$ (0 V, $-0.02$ V, $-0.04$ V) at fixed values of $V_{G1} = -0.82$ V, $V_{G5} = -0.2$ V and $V_{SD} = 0.5$ mV. (f) Stability diagram of the DQD at $V_{SD} = 2$ mV with barrier gate voltages of $V_{G1} = -0.82$ V, $V_{G3} = 0.008$ V and $V_{G5} = -0.17$ V.

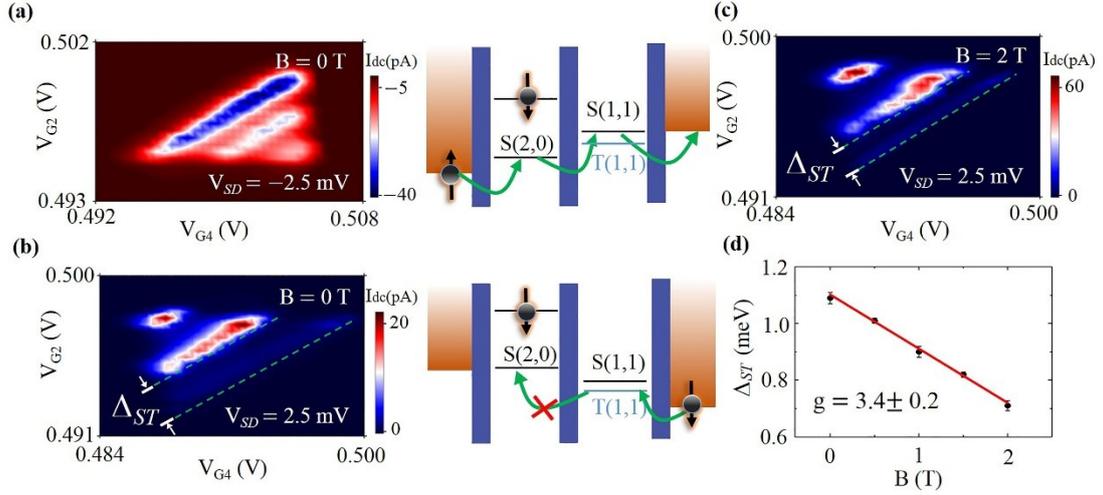

**Fig. 2.** (a) Source-drain current measured for the DQD (second device) as a function of the gate voltages $V_{G2}$ and $V_{G4}$ at $V_{SD} = -2.5$ mV and B = 0 T. The schematic shows that holes transport through the DQD via an interdot transition. (b) At a positive bias $V_{SD} = 2.5$ mV, the current is suppressed inside the singlet-triplet gap $\Delta_{ST}$ (green dashed line). The schematic presents the PSB for a hole DQD. (c) The same as in (b) but for a magnetic field of B = 2 T. (d) Singlet-triplet splitting energy $\Delta_{ST}$ of the DQD as a function of the magnetic field. A linear fit (the red line) yields g.

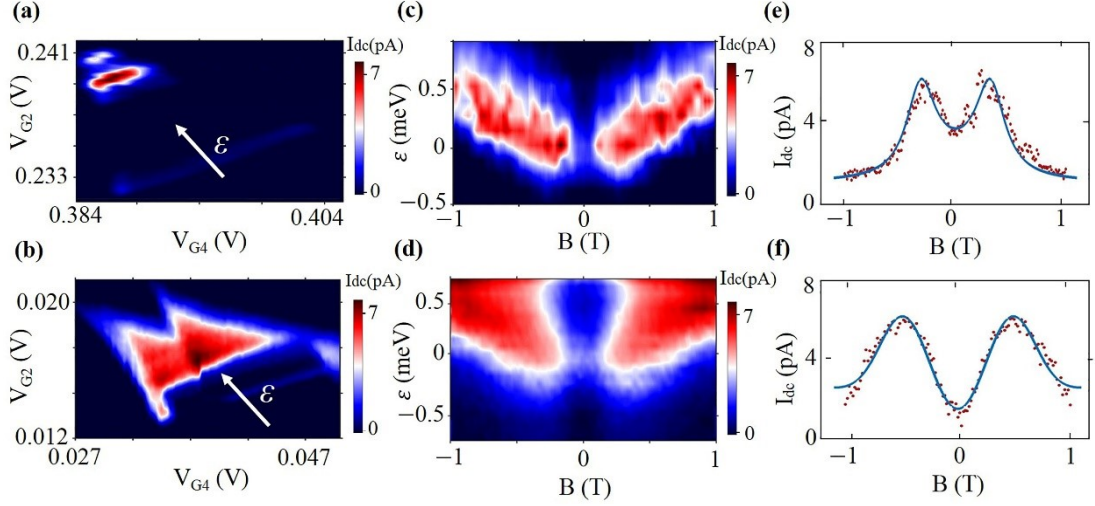

**Fig. 3.** (a) The bias triangles with the spin blockade at $V_{SD} = 2.5$ mV. (b) Spin blockade region at the same barrier gate voltages as in (a), but with more holes in the DQD. (c) and (d) The leakage current as a function of the detuning ε [white arrow in (a) and (b)] and out-of-plane magnetic field B. (e) and (f) show line cuts at ε = 0 in panels (c) and (d). The theory curves are fitted (blue lines) assuming that the PSB is spin-orbit-mediated.

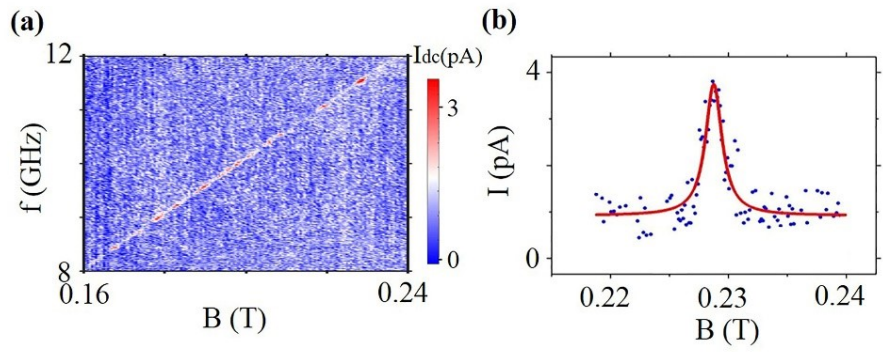

**Fig. 4.** (a) The leakage current as a function of magnetic field and microwave frequency. EDSR peak position shows linear relation between the magnetic field and microwave frequency. (b) Fitting to the resonance peaks at the microwave frequency ~ 11.5 GHz.